\documentclass[reprint,superscriptaddress,
amsmath,amssymb,
aps,
pra
]{revtex4-2}

\usepackage{graphicx}
\usepackage{color}
\usepackage[colorlinks=true,citecolor=blue,urlcolor=black]{hyperref}
\usepackage{mathrsfs}   
\usepackage{booktabs}   
\usepackage{amsthm}     

\newtheorem{definition}{Definition}

\begin{document}

\title{Provably-secure randomness generation from \\switching probability of magnetic tunnel junctions}

\author{Hong Jie Ng}
\thanks{These two authors contributed equally}
\affiliation{Department of Electrical \& Computer Engineering, National University of Singapore, Singapore }

\author{Shuhan Yang}
\thanks{These two authors contributed equally}
\affiliation{Department of Electrical \& Computer Engineering, National University of Singapore, Singapore }

\author{Zhaoyang Yao}
\affiliation{Department of Electrical \& Computer Engineering, National University of Singapore, Singapore }

\author{Hyunsoo Yang}
\email{eleyang@nus.edu.sg}
\affiliation{Department of Electrical \& Computer Engineering, National University of Singapore, Singapore }

\author{Charles C.-W. Lim}
\email{charles.lim@nus.edu.sg}
\affiliation{Department of Electrical \& Computer Engineering, National University of Singapore, Singapore }

\begin{abstract}
    In recent years, true random number generators (TRNGs) based on magnetic tunnelling junction (MTJ) have become increasingly attractive. This is because MTJ-based TRNGs offer some advantages over traditional CMOS-based TRNGs, such as smaller area and simpler structure. However, there has been no work thus far that quantified the quality of the raw output of an MTJ-based TRNG and performed suitable randomness extraction to produce provably-secure random bits, unlike their CMOS-based counterparts. In this work, we implement an MTJ-based TRNG and characterise the entropy of the raw output. Using this information, we perform post-processing to extract a set of random bits which are provably-secure.
\end{abstract}

\maketitle

\section{Introduction}

Random numbers are used in a myriad of applications, from Monte Carlo simulations for scientific research to encryption for data security. Without a doubt, random numbers are widely sought after, and much research has been done to construct random number generators (RNGs) that output high quality random numbers. The quality of random numbers can be characterised by two properties - uniformity and predictability. Indeed, in the ideal scenario, random numbers ought to be uniformly random and unpredictable from any perspective.

There are typically two classes of generators. They are the \textit{pseudo random number generator} (PRNG) and \textit{true random number generator} (TRNG). PRNGs make use of an algorithm, together with a short input random seed, to produce a longer sequence of random numbers that are uniformly distributed. However, PRNGs are deterministic algorithms at their core. Thus, any party with knowledge of the seed could predict the output of the PRNG perfectly. It is also possible for an observer to use the first few outputs of a PRNG to gain information about its current state, allowing the observer to accurately predict the subsequent outputs of the PRNG. RandCrack \cite{maxim_randcrack_2017}, a Python script that was written in 2017, only requires the first 624 numbers of the Mersenne Twister \cite{matsumoto_mersenne_1998} PRNG to be able to predict the following output numbers with high accuracy. In fact, the Mersenne Twister PRNG is widely used in C, Matlab, and Python due to its supposedly strong generator properties.

Briefly speaking, TRNGs utilise the randomness from physical processes (e.g. time taken for atomic decay) to generate random numbers. These processes are inherently random and unpredictable. Hence, they do not exhibit the shortcomings of PRNGs. Over the years, there have been many variations of TRNGs that have been implemented \cite{petrie_noise-based_2000, brederlow_low-power_2006, liu_true_2011, tang_true_2014, mathew_mu_2016, kim_82-nw_2017, chandrasekaran_036-mw_2020}. TRNGs based on magnetic tunnelling junction (MTJ) were first proposed in 2014 \cite{won_ho_choi_magnetic_2014}. Compared to their CMOS-based counterparts, some attractive features of MTJ-based TRNGs include a smaller area, a simpler structure, and scalability with silicon large-scale integration technology \cite{fu_overview_2021}. 

\begin{figure}
    \centering
    \includegraphics[width=0.45\textwidth]{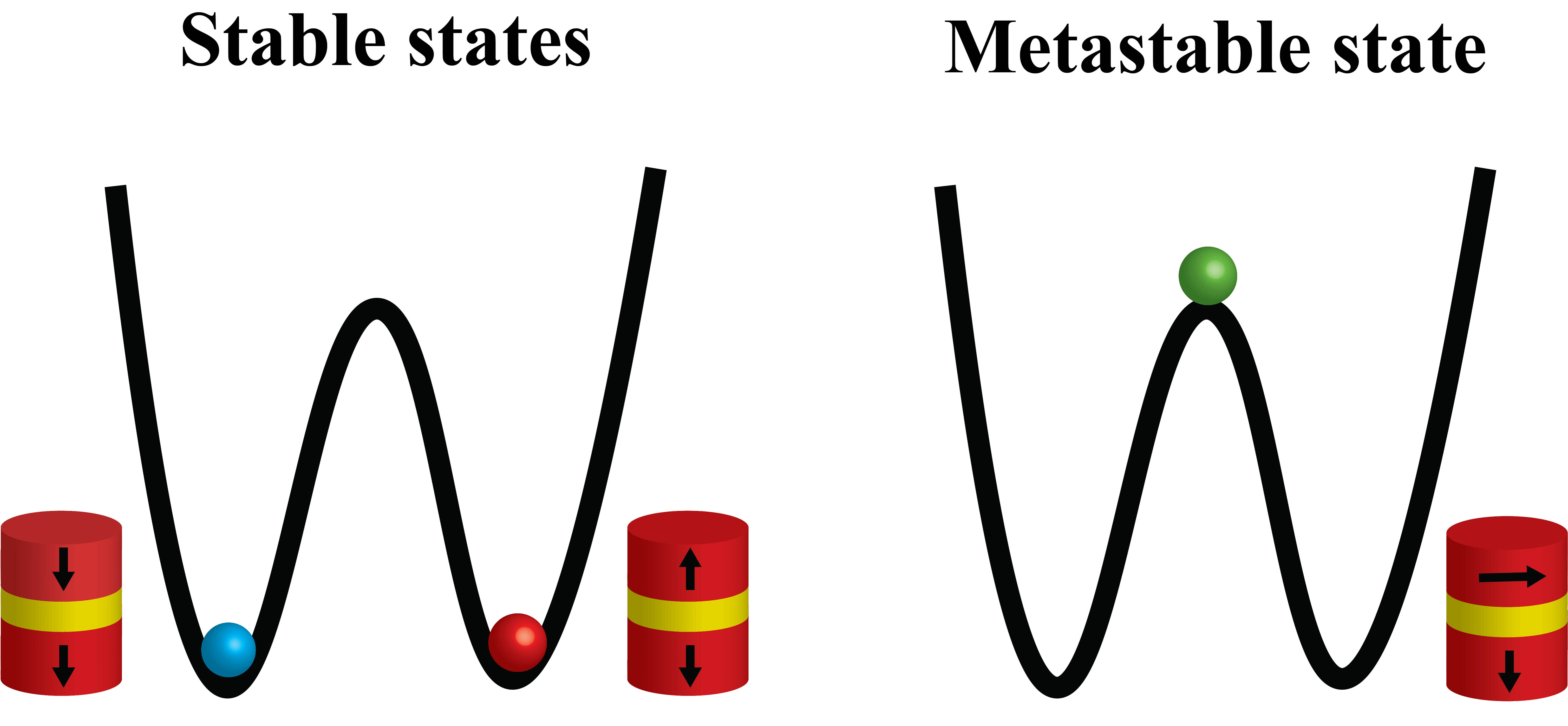}
    \caption{Illustration of STT-MTJ random number generation principle. (left) Two stable low energy states; (right) the metastable state.}
    \label{fig:MTJ_perturb}
\end{figure}

An MTJ device consists of two magnetic layers, which are the fixed layer and free layer. Depending on the relative magnetic alignment of the two magnetic layers, the MTJ exhibits two different stable states, namely the parallel (P) state and anti-parallel (AP) state. To switch between these states, we can take advantage of spin transfer torque (STT), which utilizes current to manipulate the direction of the magnetic moment of the free layer \cite{brataas2012current}. In a memory device, current pulses are applied to set the MTJ in one of the stable states. However, by adjusting the amplitude and width of the current pulses, the MTJ can be perturbed to the metastable state, as shown in Fig. \ref{fig:MTJ_perturb}. At the metastable state, thermal fluctuations determine the final state of the MTJ and give rise to randomness \cite{won_ho_choi_magnetic_2014}.

A system is considered as \textit{provably-secure} if the system's security can be formally expressed or bounded using mathematical statements, under some general assumptions about the generator and its physics. This is in contrast to the usual heuristic analysis, which typically considers the bounded adversary model, i.e., computationally bounded adversaries. Indeed, a provably-secure framework is most desirable, especially for important systems such as encryption and random number generation, where their security has to be lasting and independent of technological advancements. In addition, such a classification provides an operational meaning as well. In the case of RNGs, this definition implies that an adversary will be unable to distinguish the output of the RNG from a truly uniform and random data, except with a vanishingly small probability.

To the best of our knowledge, previous works that have implemented MTJ-based TRNGs did not obtain provably-secure random numbers. One of the works \cite{yang_28nm_2018} performed entropy estimation on a 1 Mbit dataset, which is too small in size. Thus, they did not perform suitable post-processing to obtain provably secure random numbers. 
Other works did not characterise the entropy of the raw output \cite{won_ho_choi_magnetic_2014} or provided only a partial characterisation of the raw output by assuming that the raw bits are independent \cite{vatajelu_high-entropy_2019, perach_asynchronous_2019}. In this work, we implement a MTJ-based TRNG setup and characterise the min-entropy of the raw output. Our min-entropy characterisation is carried out with minimal assumptions as we do not assume that the raw output bits are independent. We then perform suitable post-processing in the form of randomness extraction to obtain a set of provably-secure random numbers.


The remaining of this paper is organised as follows. In section \ref{implementation}, we provide the implementation details for our work, including the experimental setup of our MTJ-based TRNG and the tuning of parameters. In section \ref{ExtractionTheory}, we explain the idea behind provably-secure randomness extraction, and give a mathematical definition for the ideal output of an RNG. In section \ref{extraction}, we characterise the raw output of our MTJ-based TRNG and choose a suitable function to perform randomness extraction. Lastly, we present the final post-processed results in section \ref{results} and give a conclusion in section \ref{discussion}.

\section{Implementation} \label{implementation}


\begin{figure}[!t]
    \centering
    \includegraphics[width=0.45\textwidth]{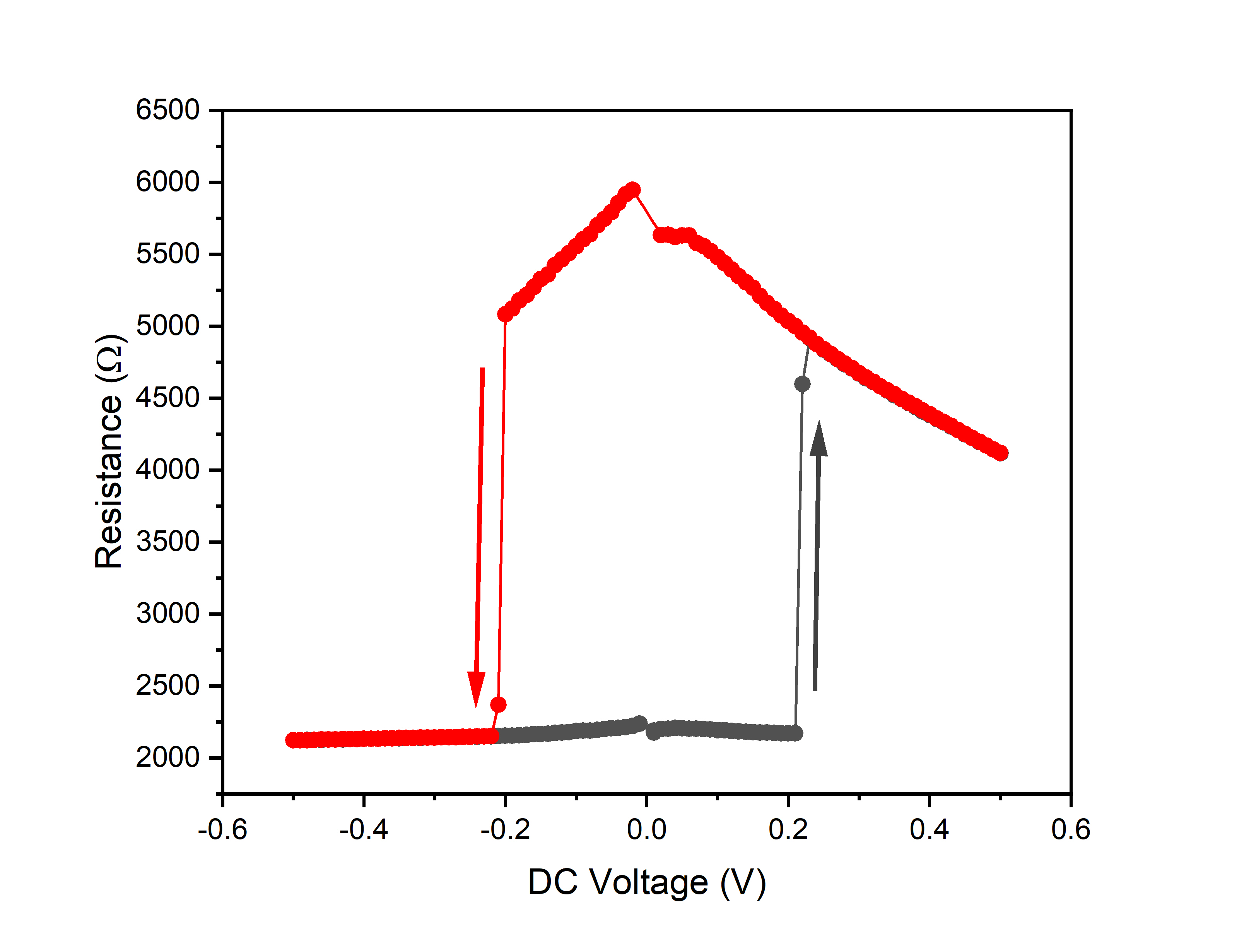}
    \caption{Measured R-V hysteresis curve of the fabricated MTJ device. The gray arrow represents the P to AP switching, and the red arrow represents the AP to P switching. }
    \label{fig:hysteresis}
\end{figure}

In this work, the TRNG is implemented using STT-MTJ. Fig. \ref{fig:hysteresis} shows the measured resistance against voltage (R-V) hysteresis curve of the fabricated MTJ device. The tunnelling magneto-resistance ratio (TMR) is $\sim$200\%. The writing operation of the STT-MTJ is a stochastic process, which is influenced by thermal fluctuations \cite{SLONCZEWSKI1996L1,PhysRevB.54.9353, doi:10.1063/1.3637545}. 
By carefully tuning the input pulses' amplitude and width, the switching probability of MTJ can be set to 50\%. This makes it possible to implement TRNG using STT-MTJ. Moreover, because MTJ-based TRNGs can be integrated with CMOS chips in high density \cite{Fukushima_2014}, it can be a great candidate for future true random number generation devices. 

\begin{figure}[!t]
    \centering
    \includegraphics[width=0.45\textwidth]{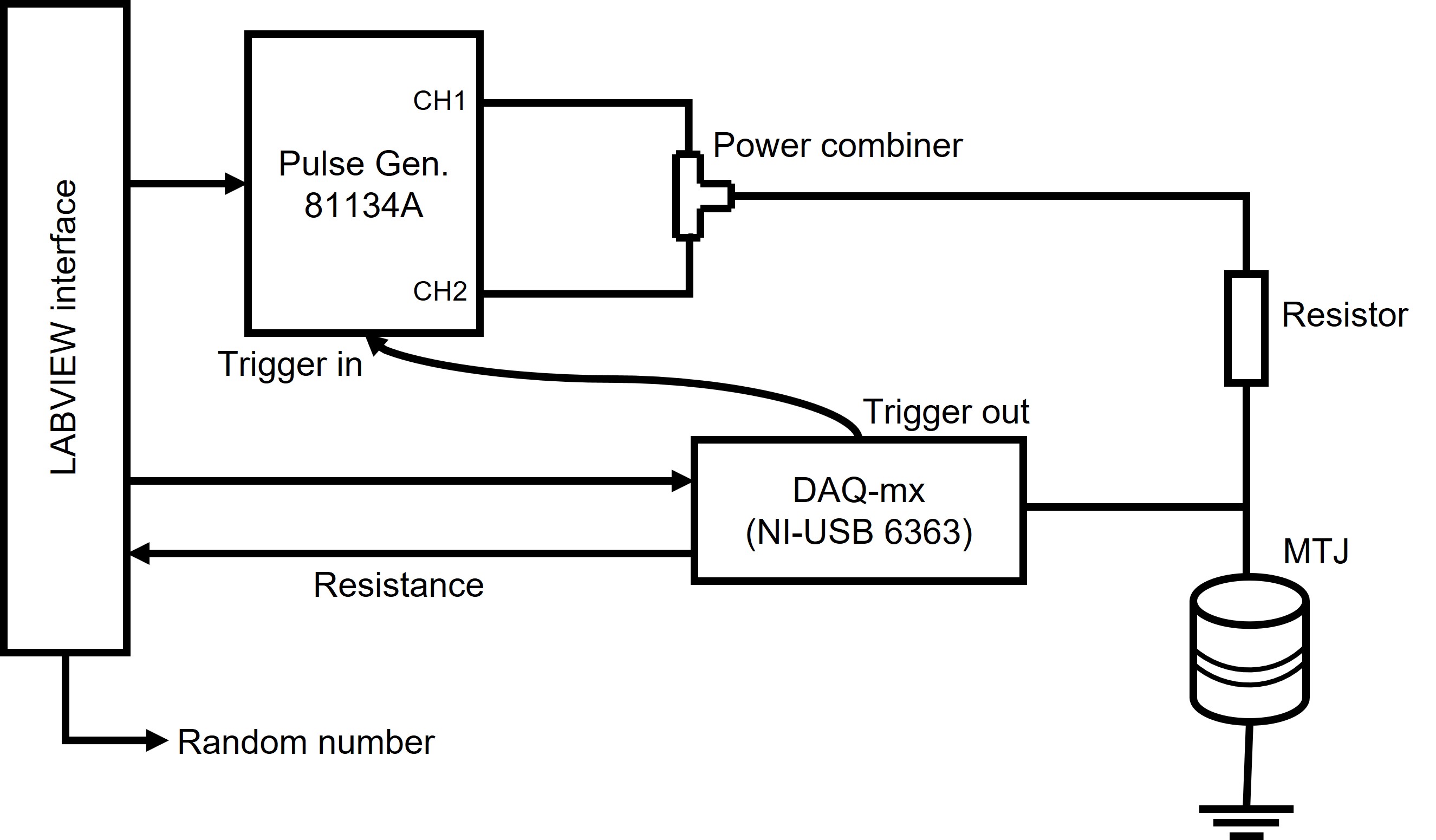}
    \caption{Random number generation setup. Keysight 81134A generates reset pulses from CH1, and perturb pulses from CH2. The trigger signal from the DAQ card ensures the simultaneous start of the pulse generation and the reading operation. Equipment is controlled by LabVIEW interface.}
    \label{fig:setup}
\end{figure}

Fig. \ref{fig:setup} shows the setup of our MTJ-based TRNG. The setup includes a pulse generator (Keysight 81134A) for generating reset and perturb pulses, a DAQ-mx (NI-USB 6363), and a R=3 k$\Omega$ resistor for reading operation. To guarantee the simultaneous start of the pulse generation and the reading operation, a trigger signal from DAQ to the pulse generator is applied. The negative reset pulse from CH1 of the pulse generator and the positive perturb pulse from CH2 are combined in the power combiner. Due to the non-volatility of MTJ devices, a reset step is required before every bit of random number generation. Thus, a random number generation cycle consists of three steps: reset, perturb, and read \cite{Fukushima_2014}. 

\begin{figure}[!t]
    \centering
    \includegraphics[width=0.45\textwidth]{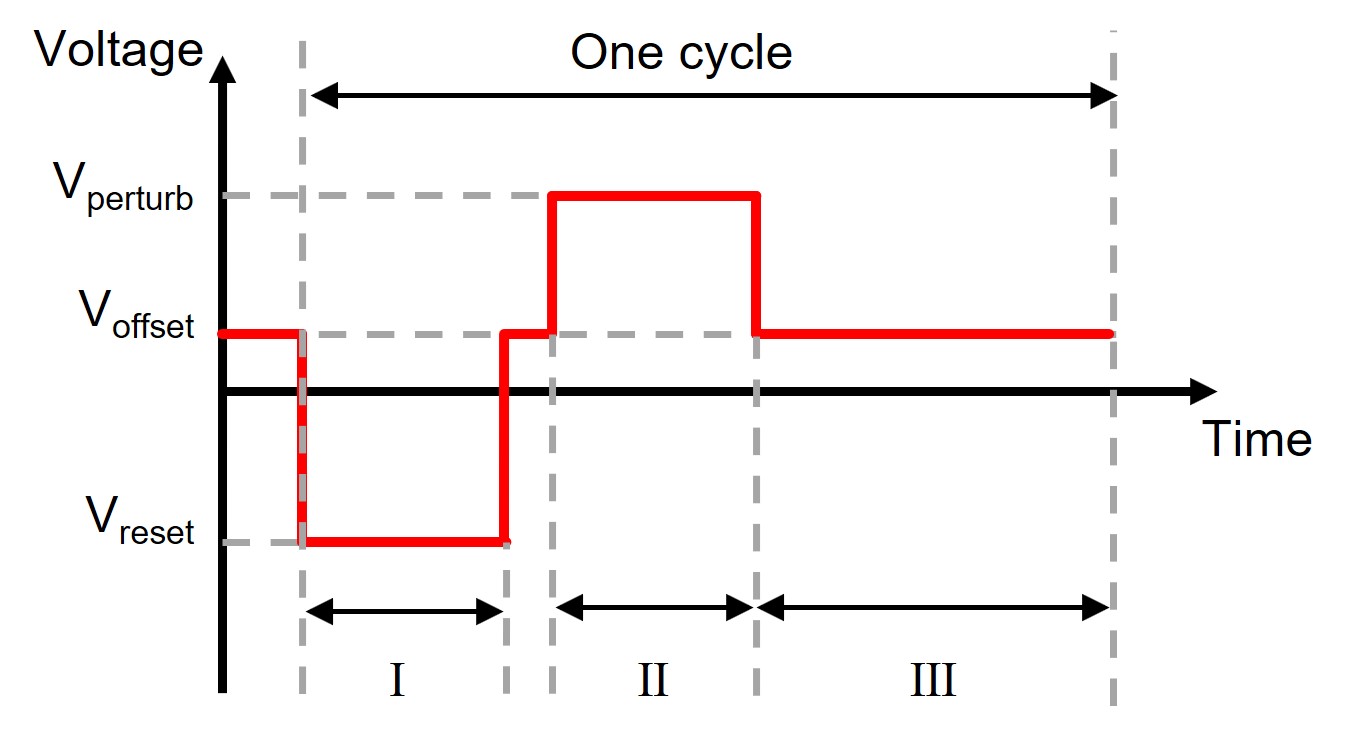}
    \caption{Random number generation time sequence. One cycle of random number generation comprises three sub-sequences, which are \uppercase\expandafter{\romannumeral1}: reset operation; \uppercase\expandafter{\romannumeral2}: perturb operation;
    \uppercase\expandafter{\romannumeral3}: reading operation.}
    \label{fig:cycle}
\end{figure}

For the reading purpose, a DC offset voltage ($V_{\text{offset}}$) of 50 mV is superpositioned to the input signal. The MTJ resistance can be expressed as
\begin{equation}
    R_{\text{MTJ}} = \frac{V_{\text{MTJ}}}{V_{\text{offset}} - V_{\text{MTJ}}} R.
\end{equation}
where  $R_{\text{MTJ}}$ is the MTJ resistance, $V_{\text{MTJ}}$ is the voltage drop across the MTJ, $V_{\text{offset}}$ is the offset voltage applied for reading purpose, R is the resistor in series to the MTJ. 
Fig. \ref{fig:cycle} shows the time sequence of one random bit generation cycle. A negative reset pulse ($V_{\text{reset}}$) with an amplitude of -900 mV and width of 1 $\mu$s is first applied to reset the MTJ to the low resistance parallel state. After the reset pulse, a positive pulse ($V_{\text{perturb}}$) with an amplitude of 652 mV and width of 1 $\mu$s is applied to perturb the MTJ to the metastable state. One cycle of random number generation takes 8 $\mu$s. The reading operation performed by the DAQ card is conducted at 2 MHz throughout the whole process. Thus, the DAQ card can read out 16 data points for one cycle. To obtain MTJ resistance, only data points corresponding to sequence \uppercase\expandafter{\romannumeral3} should be used. To ensure this, it is required that the data points corresponding to reset and perturb pulses can be clearly identified. Therefore, the pulse width of reset and perturb pulses cannot be too short, otherwise it would be hard to differentiate them. This is why in our case, 1 $\mu$s pulse width is chosen. After selecting out the correct sequence of data points, we take the average, and set 4 k$\Omega$ as the threshold 
in order to split the raw data into 2 bins. Finally, binary random numbers are output according to the final state in which the MTJ is in.


\section{Provably-secure randomness extraction from weakly random source} \label{ExtractionTheory}

Consider an RNG that outputs a binary string. In the ideal scenario, the output has to satisfy 2 conditions, which are 1) all the bits in the string are uniformly distributed, and 2) all the bits are independent from any perspective. An $m$-bit string that satisfies both criteria is referred to as the ideal output and denoted as $U_m$. 

For our MTJ-based TRNG, the raw output is most definitely not uniformly distributed. This is because the thermal noise (or any of the other noise sources present) is unlikely to remain constant throughout the entire experiment. Deviations in the thermal noise, no matter how small in magnitude, would skew the distribution of the output bits away from the uniform distribution. In addition, it would be naive to assume that the raw output bits are independent of each other. Imperfections in the experimental setup (e.g. noisy reader) might introduce correlations to the raw output bits. In fact, the raw output data failed the permutation test and Chi-square test outlined in the NIST 800-90B test suite \cite{turan_nist_2018}, which tests for independence. A sequence under test can be classified as independent and identically distributed (i.i.d.) if they pass these tests.


Clearly, the $n$-bit raw output of our MTJ-based TRNG does not satisfy the 2 criteria stated at the beginning of the section. Hence, it is classified as a weakly random source, which we denote as $X$. We thus have to utilise functions known as randomness extractors (with the help of a seed, $S$) to extract an $m$-bit almost ideal output, $Z$, from the raw output of our MTJ-based TRNG. The process can be written as $Z = \mathrm{Ext}(X,S)$. We use the notion of statistical distance to determine how close $Z$ is to the ideal output $U_m$. 

The statistical distance between two random variables, $Y$ and $Y'$, is defined as 
\begin{equation}
    \triangle(Y, Y') := \frac{1}{2}\sum_{y \in \mathscr{Y}} \bigg|P_Y(y) - P_{Y'}(y) \bigg|,
\end{equation}
where $\mathscr{Y}$ is the range of values that $Y \text{ and } Y'$ can take. If $Y$ and $Y'$ are identically distributed, then their statistical distance is 0. On the other hand, if they are perfectly distinguishable, then their statistical distance is 1. 

Intuitively, $m<n$, because it is not possible to generate randomness from nothing. This also means that for the extraction to be possible, $X$ must contain at least $m$-bits of randomness. To characterise the amount of extractable random bits that is present in a weakly random source, we use the notion of min-entropy. The min-entropy of $X$ is defined as
\begin{equation}
    H_{\mathrm{min}}(X) := \min_{x \in \mathscr{X}} \bigg[- \mathrm{log}_2 P_X(x) \bigg],
\end{equation}
where, $\mathscr{X}$ is the range of values that $X$ can take. Notice that we use min-entropy instead of the Shannon entropy to characterise the extractable randomness. To explain why, let us consider an extractor that outputs an ideal $Z$, i.e. $P_Z(z) = 2^{-m} \text{ for all } z \in \mathscr{Z}$, where $\mathscr{Z}$ is the range of values that $Z$ can take. Any $z \in \mathscr{Z}$ thus has to satisfy 
\begin{equation}
    P_Z(z) = \sum_{x: \mathrm{Ext}(X) = z} P_X(x) = 2^{-m}.
\end{equation}
Clearly, the implication is that $P_X(x) \leq 2^{-m}$ for all $x \in \mathscr{X}$. Therefore, to extract $m$ random bits from $X$, it is necessary that $H_{\mathrm{min}}(X) \geq m$.

Now that the preliminaries have been introduced, we provide a formal definition for randomness extraction. 
\begin{definition}[Strong seeded extractors \cite{nisan_randomness_1996}]
The function $\mathrm{Ext} : \{ 0,1 \}^n \times \{ 0,1 \}^r \rightarrow \{ 0,1 \}^m$ is a $(k,\varepsilon)$-strong extractor if for any source $X$ with $H_{\mathrm{min}}(X) \geq k$ and seed $S = U_r$, its output $Z = \mathrm{Ext}(X,S)$ satisfies $\triangle(ZU_r, U_m U_r) \leq \varepsilon$, where $YY'$ represents the concatenation of the strings $Y$ and $Y'$. 
\end{definition}

The implication of a strong extractor is that the output $Z$ remains uniformly random even in the presence of seed $S$, i.e. $Z$ and $S$ are independent of each other. Operationally, this means that we are able to concatenate the seed used to the output of the strong extractor, without any increase in the security parameter. This is important because in Toeplitz hashing, the seed used is longer than the output length. Thus, if we do not use a strong extractor, there will be no net randomness extracted since we consume more randomness than we extract. We follow up by providing the definition of universal hashing.

\begin{definition}[Universal hash functions \cite{carter_universal_1979}]
A family of hash functions $\mathscr{F}=\{ f : \{ 0,1 \}^n \rightarrow \{ 0,1 \}^m \} \text{ with } n \geq m$ is called universal if for $f$ randomly chosen from $\mathscr{F}$,
\begin{equation}
    \mathrm{Pr}[f(x) = f(x')] \leq 2^{-m}, \hspace{2mm} \forall \text{  } x \neq x' \in \{ 0,1 \}^n.
\end{equation}
\end{definition}
In our application, we use the random seed $S$ to choose a function from the family of universal hash functions. We denote the chosen function as $f_s$. This also means that the family of universal hash functions has size $|\mathscr{F}| = 2^r$. Importantly, the work in \cite{impagliazzo_pseudo-random_1989} showed that universal hash functions can be used to construct strong seeded extractors. We provide the definition below. 

\begin{definition}[Leftover hash lemma \cite{impagliazzo_pseudo-random_1989}]
Let $X$ be a min-entropy source with $H_{\min}(X) \geq k$ and $\mathscr{F} = \{ f_s : \{ 0,1 \}^n \rightarrow \{ 0,1 \}^m \}$ be a universal hashing family of size $2^r$ with $m = \lfloor k - 2\log_2 (1/\varepsilon) + 2 \rfloor$. Furthermore, let $\mathrm{Ext}(X,s)=f_s(X)$. Then, the extractor is a $(k,\varepsilon)$-strong extractor with seed length $r$ and output length $m$.
\end{definition}

Hence, using a family of universal hash functions to perform randomness extraction allows us to obtain an extracted output that is provably-secure, i.e. we are able to bound how close our extracted output is to the ideal output. Another advantage is that such a security definition also adheres to the universal composability framework proposed in \cite{canetti_universally_2001, pfitzmann_composition_2000}. The implication is that if we use the extracted output in other protocols, the security parameter of that protocol will increase by $\varepsilon$.

\section{Entropy estimation and extraction} \label{extraction}

We shall denote the $n$-bit raw binary random numbers that our MTJ-TRNG produces as $X$. We can classify $X$ as uniformly random if $H_{\text{min}}(X) = n$. However, as mentioned in section \ref{ExtractionTheory}, more often than not, our MTJ-TRNG produces a string that is not uniformly random, i.e. $H_{\text{min}}(X) < n$. Thus, we have to lower-bound the min-entropy of $X$ using techniques like the NIST 800-90B entropy estimation suite \cite{turan_nist_2018} to certify that $H_{\text{min}}(X) \geq k$. Using this estimate of $k$, we are able to quantify the amount of extractable randomness in $X$ and extract an $m$-bit output string $Z$, such that $m \leq k$. The string produced is $\varepsilon$-close to the ideal uniform $m$-bit distribution, $\text{U}_m$, i.e. $\triangle (Z,\text{U}_m) \leq \varepsilon$.

The results that we obtain from NIST 800-90B indicate that the min-entropy per bit of $X$ is 0.778584. It should be noted that we do not assume independence of the raw bits when performing min-entropy estimation. We collected $n=217,504,350$ bits, which translates to $k=169,345,406$ bits. We set our security parameter to be $\varepsilon = 10^{-10}$, and employ the leftover hash lemma \cite{impagliazzo_pseudo-random_1989} to obtain $m = \lfloor k - 2\text{log}_2(1/\varepsilon) + 2 \rfloor = 169,345,260$ bits.

The extraction is carried out via Toeplitz hashing, which belongs to a family of universal hash functions \cite{krawczyk_lfsr-based_1994}. Toeplitz hashing utilises a Toeplitz matrix, $\textbf{T}$. The Toeplitz matrix is an $m \times n$ diagonal-constant matrix, and it is constructed by filling up the first column and first row of the matrix with a uniform seed, denoted by $S$. Thus, the seed length required for Toeplitz hashing is $n+m-1$ bits. $Z$ is obtained by performing $Z = \textbf{T} \cdot X$, which is expressed as
\begin{equation*}
    \begin{bmatrix}
    s_n & s_{n-1} & \cdots &  s_2 & s_1\\
    s_{n+1} & s_n & \cdots & s_3 & s_2\\
    \vdots & \vdots & \ddots &  \vdots & \vdots\\
    s_{n+m-1} & s_{n+m-2} & \cdots & s_{m+1} & s_m
    \end{bmatrix} \cdot 
    \begin{bmatrix}
    x_{1} \\
    x_{2} \\ 
    \vdots \\ 
    x_{n-1} \\
    x_{n}
    \end{bmatrix} = \begin{bmatrix}
    z_1 \\
    z_2 \\
    \vdots \\
    z_m
    \end{bmatrix}.
\end{equation*}

The computational complexity of Toeplitz hashing is $O(n^2)$ due to the need to perform matrix-vector multiplication. However, this is undesirable, as $n$ needs to be a large number for us to be able to accurately characterise the min-entropy of the raw data. To improve the computational complexity of Toeplitz hashing, we utilise the fast Fourier transform (FFT) algorithm. The FFT algorithm speeds up matrix-vector multiplications when the matrix used is a circulant matrix. In this case, Toeplitz matrix \textbf{T} is not a circulant matrix, but it can easily be converted into one by padding \textbf{T} with additional rows and columns, and filling up these rows and columns with 0s and 1s accordingly to obtain $\textbf{T}_{\text{circ}}$. As $\textbf{T}_{\text{circ}}$ has a larger number of columns than \textbf{T}, we have to pad $X$ with 0s to get $X_{\text{pad}}$. The main process of Toeplitz hashing accelerated by FFT can then be expressed as
\begin{equation*}
    Z_{\text{pad}} = IFFT(FFT(X_{\text{pad}}) \cdot FFT(\text{T'}_{\text{circ}})) = \textbf{T}_{\text{circ}} \cdot X_{\text{pad}},
\end{equation*}
where $FFT \text{ and } IFFT$ are the FFT algorithm and its inverse respectively, $\text{T'}_{\text{circ}}$ is the first column of $\textbf{T}_{\text{circ}}$, and $Z_{\text{pad}}$ is the extended output. The original output $Z$ can be recovered by keeping the first $m$ bits of $Z_{pad}$. The computational complexity of Toeplitz hashing with FFT is $O(n \log n)$, which is an exponential improvement from the initial computational complexity.

\section{Results} \label{results}


\begin{table*}[!t]
    \centering
    \setlength{\tabcolsep}{0.5em}
    \begin{tabular}{l c c c c c c}
    \toprule
    {} & \multicolumn{3}{c}{\textbf{Raw data}} & \multicolumn{3}{c}{\textbf{Extracted data}}  \\
    \cmidrule(lr){2-4} \cmidrule(lr){5-7}
    \textbf{Statistical test} & \textbf{P-value} & \textbf{Proportion} & \textbf{Result} & \textbf{P-value} & \textbf{Proportion} & \textbf{Result} \\
    \midrule
    Frequency & 0.0000 & 0.1901 & Failure & 0.2315 & 1.0000 & Success \\
    Block frequency & 0.0000 & 0.8651 & Failure & 0.9233 & 0.9882 & Success \\
    Cumulative sums & 0.0000 & 0.1839 & Failure & 0.4201 & 1.0000 & Success \\
    Runs & - & - & - & 0.8606 & 0.9941 & Success \\
    Longest run & 0.0000 & 0.1797 & Failure & 0.6436 & 1.0000 & Success \\
    Rank & 0.1806 & 1.0000 & Success & 0.4547 & 0.9822 & Success \\
    FFT & 0.0000 & 0.5253 & Failure & 0.4666 & 1.0000 & Success \\
    Non-overlapping template & 0.0000 & 0.0000 & Failure & 0.0001 & 0.9704 & Success \\
    Overlapping template & 0.0000 & 0.0335 & Failure & 0.0024 & 0.9941 & Success \\
    Universal & 0.0000 & 0.0000 & Failure & 0.6825 & 0.9882 & Success \\
    Approximate entropy & - & - & - & 0.5659 & 0.9941 & Success \\
    Random excursions & 0.0003 & 0.8182 & Failure & 0.2307 & 0.9703 & Success \\
    Random excursions variant & 0.0909 & 0.9091 & Success & 0.0780 & 0.9604 & Success \\
    Serial & 0.0000 & 0.0000 & Failure & 0.1597 & 0.9763 & Success \\
    Linear complexity & 0.3781 & 0.9862 & Success & 0.3979 & 0.9882 & Success \\
    \bottomrule
    \end{tabular}
    \caption{Results of the NIST 800-22 statistical test suite for the raw and post-processed data. The minimum values for each test are tabulated and rounded down to 4 decimal places. To pass an individual statistical test, the post-processed data needs to obtain 1) P-value $\geq 0.0001$; and 2) proportion $\geq 0.9670$, except for random excursion variant test, which requires proportion $\geq 0.9603$. Similarly, the raw data needs to obtain 1) P-value $\geq 0.0001$; and 2) proportion $\geq 0.9697$, except for random excursion variant test, which requires proportion $\geq 0.9000$.}
    \label{tab:NistResult}
\end{table*}

We performed Toeplitz hashing on a personal computer with an Intel Xeon E5-2697 CPU and 128 GB of random access memory (RAM). The total time taken for the extraction was 72.285 seconds, which translates to a throughput speed of 2.3427 Mbits/s.

The output from Toeplitz hashing is saved in a text file, and NIST 800-22 statistical test suite \cite{rukhin_nist_2010} was employed to test the quality of both the raw and extracted random numbers. To run the test, the extracted (raw) data is split into 169 (217) blocks, each with length $10^6$ bits. The blocks are passed to all 15 statistical tests, which perform hypothesis testing at the 1\% significance level, and returnOr  2 values, the P-value and proportion. 

The proportion parameter is the proportion of sequences that pass the test at the selected significance level. The range of acceptable proportions is determined using the confidence interval. It is calculated as
\begin{equation*}
    \widehat{p} \pm 3 \sqrt{\frac{\widehat{p}(1-\widehat{p})}{k}},
\end{equation*}
where $\widehat{p}=1 - \alpha$ and $k$ is the number of sequences. When testing the extracted output, we substitute $\alpha = 0.01 \text{ and } k = 169$ to obtain a confidence interval of 0.9670. The interpretation of this value is that if the returned value of proportion for a specific test is below 0.9670, we conclude that our extracted output fails that test and is therefore not uniformly random. Similarly, when testing the raw output, we substitute $\alpha = 0.01 \text{ and } k = 217$ to obtain a confidence interval of 0.9697.

The P-value parameter for a specific test, denoted as $P\text{-value}_T$, is the result of the Goodness-of-Fit Distributional Test on the P-values obtained for that statistical test. The Goodness-of-Fit Distributional Test tests if the distribution of P-values obtained follows a uniform distribution. If the returned $P\text{-value}_T < 0.0001$, then the P-values failed the uniformity test, and thus, the sequence under test is considered to be non-uniformly distributed as well.

The results of the NIST test for both our raw and extracted data are displayed in Table \ref{tab:NistResult}. For tests that are executed several times (e.g. non-overlapping template), the minimum values out of all the runs are tabulated, with the values rounded down to 4 decimal places. From the results, it can be seen that our raw data fails 10 out of 13 tests. The Runs test was not performed on the raw data because it carries out the Frequency test as a prerequisite. As the Frequency test returned a failure, the Runs test will be considered as failed by default. Similarly, because the Serial test failed, the Approximate Entropy test was skipped as well. 

On the other hand, after post-processing, the P-values returned for all the tests are greater than 0.0001, and all the proportion values returned are also greater than 0.9670 as well. Hence, this shows that our extracted random numbers pass all tests in the NIST statistical test suite, and that we have managed to extract high quality random numbers from our initial weakly random source.

\section{Discussion} \label{discussion}

In summary, we implemented an MTJ-based TRNG, estimated the min-entropy of the raw output bits, and performed suitable post-processing on the raw output to obtain a set of provably-secure random numbers. To the best of our knowledge, this is the first-ever implementation of provably-secure random number generation using an MTJ device. The extracted random numbers pass all the tests in the NIST STS. Our implementation also has minimal assumptions as well, as we do not even assume that our raw data are independent or identically distributed when performing entropy estimation. 

Our work could have been improved by performing an in-depth characterisation of the MTJ device to model it using mathematical statements. By doing so, we will be able to obtain an accurate value of min-entropy, instead of having to rely on min-entropy estimation tests. We could also consider the side-information that is held by an adversary, Eve. Performing min-entropy estimation on the raw data conditioned on Eve's side-information would allow us to obtain a final extracted output that is uniformly random, even from Eve's perspective. 

\section*{Acknowledgement}
We thank Chao Wang for useful inputs and discussions. We acknowledge funding support from the National Research Foundation of Singapore (NRF) Fellowship grant (NRFF11-2019-0001) and NRF Quantum Engineering Programme 1.0 grant (QEP-P2).

\bibliography{biblio}

\end{document}